\documentclass[%
 reprint,
 amsmath,amssymb,
 aps,
floatfix
]{revtex4-1}

\usepackage{graphicx}
\usepackage{bm}
\usepackage{xcolor}
\usepackage{booktabs} 
\usepackage{siunitx}
\usepackage{amsfonts, amssymb, amsmath}
\usepackage{upgreek}
\usepackage{physics} 
\usepackage{float}

\DeclareSIUnit\angstrom{\text {\AA}}
\newcommand{\thor}{$^{229}$Th}
\newcommand{\thorm}{$^\mathrm{229m}$Th}
\newcommand{\caf}{CaF$_2$}

\newcommand{\sr}{$^{87}$Sr}

\usepackage{graphicx}
\usepackage{dcolumn}
\usepackage{bm}

\begin{document}

\preprint{APS/123-QED}

\title{Frequency reproducibility of solid-state Th-229 nuclear clocks}
\author{Tian Ooi, Jack F. Doyle, Chuankun Zhang, Jacob S. Higgins, Jun Ye}
\email{email: ye@jila.colorado.edu}
\affiliation{JILA, NIST and University of Colorado, Department of Physics, University of Colorado, Boulder, CO 80309}

\author{Kjeld Beeks, Tomas Sikorsky, Thorsten Schumm}
\affiliation{Vienna Center for Quantum Science and Technology, Atominstitut, TU Wien, 1020 Vienna, Austria}

\date{\today}

\begin{abstract}
\noindent Solid-state \thor{} nuclear clocks are set to provide new opportunities for precision metrology and fundamental physics. Taking advantage of a nuclear transition's inherent low sensitivity to its environment, orders of magnitude more emitters can be hosted in a solid-state crystal compared to current optical lattice atomic clocks. Furthermore, solid-state systems needing only simple thermal control are key to the development of field-deployable compact clocks.
In this work, we explore and characterize the frequency reproducibility of the \thor{}:CaF$_2$ nuclear clock transition, a key performance metric for all clocks.
We measure the transition linewidth and center frequency as a function of the doping concentration, temperature, and time.
We report the concentration-dependent inhomogeneous linewidth of the nuclear transition, limited by the intrinsic host crystal properties. We determine an optimal working temperature for the \thor{}:CaF$_2$ nuclear clock at \SI{195(5)}{\kelvin} where the first-order thermal sensitivity vanishes. 
This would enable \textit{in-situ} temperature co-sensing using different quadrupole-split lines, reducing the temperature-induced systematic shift below the 10$^{-18}$ fractional frequency uncertainty level. 
At 195\,K, the reproducibility of the nuclear transition frequency is 280 Hz (fractionally $1.4\times10^{-13}$) for two differently doped \thor{}:CaF$_2$ crystals over four months. These results form the foundation for understanding, controlling, and harnessing the coherent nuclear excitation of \thor{} in solid-state hosts, and for their applications in constraining temporal variations of fundamental constants.
\end{abstract}
\normalfont
\maketitle

\begin{figure*}
    \includegraphics[width=14cm]{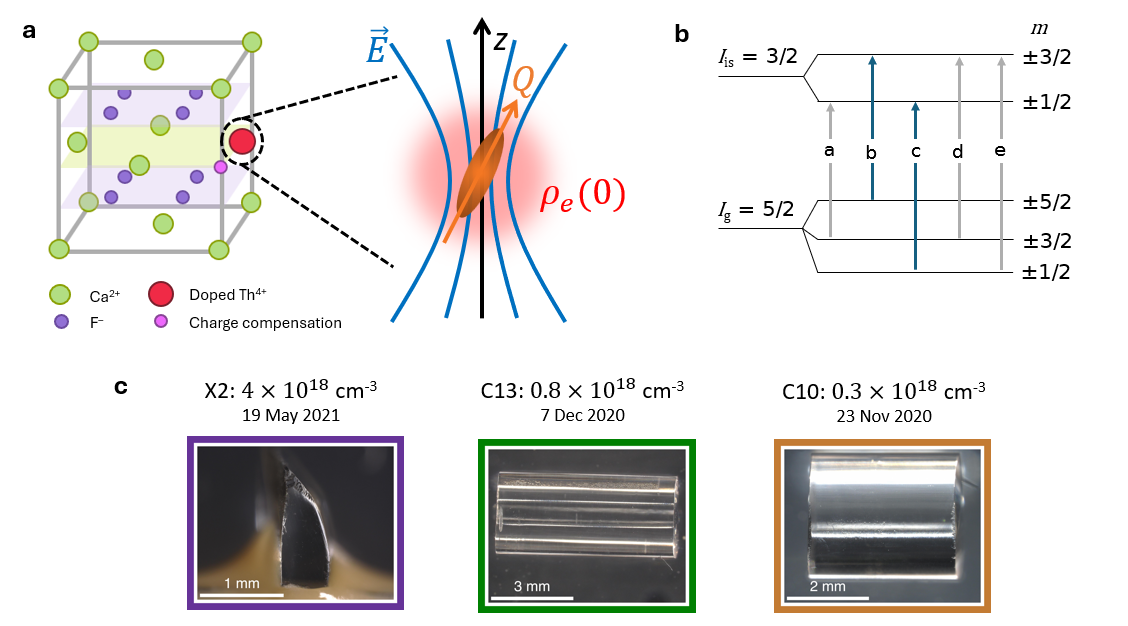}
    \caption{Context for the characterization of nuclear clock reproducibility.
    \textbf{a}, A \thor{} nucleus hosted inside a \caf{} crystal experiences a local electric field gradient $\nabla\vec{E}$ and a local charge density $\rho_\mathrm{e}(0)$. The \thor{} nuclear electric quadrupole moment $Q$ interacts with $\nabla\vec{E}$, leading to split energy levels.
    \textbf{b},  Level diagram with electric quadrupole splitting. The ground (spin $I_\mathrm{g} = 5/2$) and excited isomeric state (spin $I_\mathrm{is}=3/2)$ are split to 3 and 2 sub-levels respectively. $I$ denotes nuclear angular momentum and $m$ is the projection of $I$ along the z-axis. 
    Five allowed transitions are labeled as lines a--e. 
    Line b and c (highlighted in blue) are characterized in this work, as their properties are desirable for construction of a clock.  
    \textbf{c}, Microscope images of the three crystals used in this work, with their corresponding fabrication date and \thor{} doping concentration labeled. The colored border around the image will be used throughout this paper to indicate the corresponding crystal in all further plots.
           }
    \label{fig:setup}
\end{figure*}
     
Optical atomic clocks are the world’s most precise timekeepers and are poised to answer a wealth of fundamental questions, including searches for new physics beyond the Standard Model~\cite{ludlow_optical_2015}.  Taking advantage of electronic transitions in atoms with long coherence times, optical clocks have achieved systematic uncertainties below $10^{-18}$~\cite{aeppli_clock_2024, marshall2025high} with measurement precision approaching $10^{-21}$~\cite{bothwell_resolving_2022}. Optical lattice clocks achieve this superior precision by scaling the number of quantum absorbers to $10^{5}$, greatly reducing the quantum projection noise limit.

Nuclear clocks, based on transitions within nuclei as a frequency reference~\cite{zhang_frequency_2024}, can further scale the number of absorbers to more than $10^{15}$ by using solid-state hosts~\cite{beeks_growth_2023, beeks_optical_2024}, potentially opening up new regimes of clock stability.
The thorium-229 isotope has a uniquely low energy nuclear isomeric state around 8.4~eV above the ground state, corresponding to a wavelength of around 148.38~nm, accessible by tabletop lasers. This \thorm{} transition has been proposed~\cite{tkalya1996processes, peik_nuclear_2003} for a next generation optical clock~\cite{peik_nuclear_2021, beeks_thorium-229_2021, von2020229}.
The relative insensitivity of nuclear energy states to external perturbations compared to electronic states makes this type of clock inherently more stable, and suitable for embedding in a solid-state crystal~\cite{rellergert_constraining_2010, kazakov_performance_2012}.

Following a series of indirect and direct observations of the \thor{} nuclear transition energy~\cite{von_der_wense_direct_2016, thielking_laser_2018, seiferle_energy_2019, masuda_x-ray_2019, yamaguchi_energy_2019, sikorsky_measurement_2020, kraemer_observation_2023}, the first resonant excitation was achieved in a \thor{}-doped calcium fluoride (\thor:\caf) crystal \cite{tiedau_laser_2024}, followed by excitation in a \thor:LiSrAlF$_6$ crystal~\cite{elwell_laser_2024}. The first frequency-based measurement~\cite{zhang_frequency_2024} was achieved using a vacuum ultraviolet (VUV) frequency comb~\cite{zhang_tunable_2022} linked to the JILA $^{87}$Sr optical atomic clock. This work measured the transition frequency with kHz level precision and resolved the nuclear electric quadrupole splitting in \thor:\caf\ crystals. The precise measurement of the electric quadrupole ratio between the ground and excited states allowed the determination of the volume ratio of these two nuclear states and a new estimate of the transition sensitivity to variations of the fine structure constant $\alpha$~\cite{beeks_fine-structure_2024}. The \thor{} resonant excitation has also been observed in \thor{}F$_4$ and \thor{}O$_2$ thin films \cite{zhang_229thf4_2024, elwell_229th_2025}, an important step towards portable nuclear clocks that can be prepared inexpensively with minimal starting material. 

Developing a new clock platform requires a deep understanding of systematic effects that can perturb transition frequencies. These perturbations can arise from external electric and magnetic fields or the host environment. Building on our first characterization of temperature-induced frequency shifts~\cite{higgins_temperature_2025}, here we report the frequency reproducibility of the nuclear transition in \thor:\caf{} across a large range of doping concentrations, temperature, and temporal duration. We find that the crystal-limited transition linewidths do not change over a large temperature range, but do increase with thorium concentration due to dopant-induced inhomogeneous broadening. We report a \textit{zero-shift temperature} where the local slope of the transition frequency versus temperature curve vanishes. The different temperature sensitivities of two transition lines can be utilized for a co-thermometry stabilization scheme in future clock operation. Finally, we report that the transition frequency remains stable at the kilohertz level over the course of many months and across different crystals, with two of the crystals demonstrating sub-kHz reproducibility with a standard error of 280 Hz at the zero-shift temperature. This demonstrates the power of this platform as a frequency standard where frequency reproducibility is paramount.

\begin{figure*}
    \includegraphics[width=14cm]{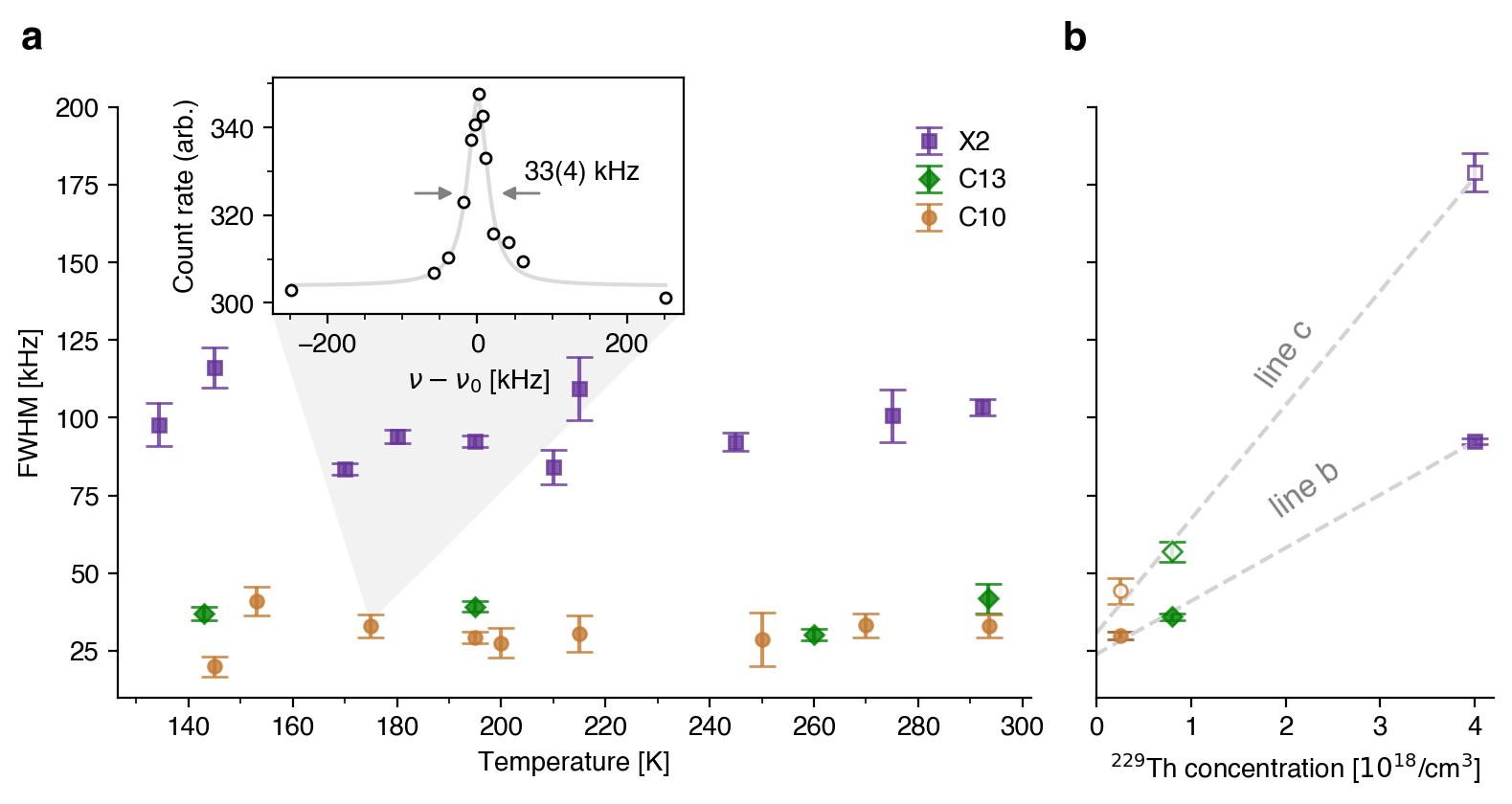}
    \caption{Characterization of the nuclear clock transition linewidth in \caf{} host. 
    \textbf{a}, Observed nuclear transition full-width-at-half-maximum (FWHM) linewidth of line b as a function of temperature across three different crystals (X2, C13, and C10). When multiple sets of data are taken at the same temperature within 1\,K, the FWHM is represented by the weighted average. The inset shows a typical line scan with data fitted to a Lorentzian line shape. The fit yields a center frequency of $\nu_0=2,020,407,298,706.6(8)$\,kHz and a linewidth of \SI{33(4)}{kHz}. For each scan, we correct the count rate and frequency by accounting for known line-skewing effects (see Methods). 
    \textbf{b}, Observed nuclear transition linewidth scaling in different crystals for both line b and line c. 
    Lower doping concentration leads to a narrower transition linewidth. 
    Linear fits to the data are presented as a visual guide. 
    Line c shows a wider linewidth compared to line b for all crystals used in this experiment. 
    }
    \label{fig:fig2}
\end{figure*}

\medskip
\noindent\textbf{State-resolved nuclear spectroscopy}
\smallskip
\par\nobreak
We use \thor:\caf\ crystals grown with the vertical gradient freeze method at TU Wien~\cite{beeks_growth_2023}.
The \thor{} atom substitutes with Ca$^{2+}$, creating a local defect  (Fig.~\ref{fig:setup}a). The exact charge compensation mechanism is not known and remains an area of active study ~\cite{takatori_characterization_2025, nalikowski_embedded_2025}.
Although the nuclear transition has reduced environmental sensitivity, a temperature-dependent frequency shift is observed, arising mainly from two effects in the host crystal~\cite{higgins_temperature_2025}. First, the \thor{} nuclear ground and excited states have different mean charge radii~\cite{beeks_fine-structure_2024, thielking_laser_2018}, leading to an electric monopole shift dependent on the local electron charge density at the nucleus $\rho_\mathrm{e}(0)$. Second, the nuclear spin and quadrupole moment also change between ground and excited state, leading to an electric quadrupole shift dependent on the local electric field gradient $\nabla\vec{E}$ (EFG). This EFG arises from the crystal and sets the quantization axis ($z$-axis in Fig.~\ref{fig:setup}a) of the nuclear spin orientation. The electric quadrupole interaction Hamiltonian~\cite{dunlap_mossbauer_1985}
$H= \frac{QV_{zz}}{4I(2I-1)} \left[ 3I_{z}^{2} - \textbf{I}^{2} + \eta\left( I_{x}^{2} - I_{y}^{2} \right) \right]$
also leads to the splitting of the nuclear states and five asymptotically allowed transitions~\cite{beeks_fine-structure_2024} given by the selection rule $\Delta m = 0, \pm1$ (Fig.~\ref{fig:setup}b).
Here, $Q$ is the spectroscopic nuclear electric quadrupole moment, $I$ is the nuclear angular momentum, and $m$ is the projection of $I$ along the z-axis. In the diagonal EFG matrix $V_{ij}$, $V_{zz}$ describes the z component and $\eta=(V_{xx}-V_{yy})/V_{zz}$ describes the field asymmetry.
As the crystal temperature and hence the lattice spacing vary, changes in $\rho_\mathrm{e}(0)$ and $\nabla\vec{E}$ cause the nuclear transitions to shift in frequency. In previous work, we measured the state-resolved transition line frequencies~\cite{zhang_frequency_2024} and characterized their temperature dependence~\cite{higgins_temperature_2025} in one \thor:\caf{} crystal, named X2~\cite{beeks_growth_2023}.

We now focus on the lines with the smallest ($m=\pm 5/2 \rightarrow m=\pm3/2$, line b) and largest ($m = \pm 1/2\rightarrow m=\pm 1/2$, line c) temperature sensitivity, highlighted in Fig.~\ref{fig:setup}b. 
We extend the nuclear spectroscopy to three different \caf{} crystals (X2, C13, and C10), with their \thor{} concentration and fabrication date listed in Fig.~\ref{fig:setup}c. X2 is glued with Epotek H77 to a MgF$_2$ disk which is in turn clamped to a copper mount, while C10 and C13 are tied with wire to a copper mount, using indium foil for improved thermal conduction.
The temperature is measured using a sensor mounted to the crystal holder and stabilized by feedback onto a heater on the copper thermal link connecting a liquid nitrogen dewar and the crystal mount~\cite{higgins_temperature_2025}. During the scans, the crystal temperature does not deviate by more than 0.1\,K. 



A typical line scan is shown in the inset of Fig. \ref{fig:fig2}a, corresponding to line b of crystal C10 at \SI{175}{K}. For each point of the scan (open circles), we irradiate the target with our VUV frequency comb for 400\,s. We then turn off the laser and count VUV photons emitted from the crystal over a 200\,s detection window. We take the average count rate over the detection window after correcting for known systematics such as fluctuations in laser intensity and residual excitation from previous points (see Methods). Fitting a Lorentzian lineshape (grey line) to the data, we extract the center frequency, full-width-at-half-maximum (FWHM) linewidth, and their uncertainties. 

Since May 2024, we have accumulated a database of 48 line scans at temperatures ranging from 134\,K to 294\,K across the three crystals. In the following, we characterize the reproducibility of the nuclear clock transition frequency and linewidth as a function of temperature, \thor{} doping concentration, and time.

\medskip
\noindent\textbf{Linewidth characterization}
\smallskip
\par\nobreak
The observed isomeric state life time of 641(4)\,s \cite{zhang_frequency_2024} would correspond to a superior transition quality factor exceeding 10$^{19}$. 
However, in prior work we observed linewidths between 200\,kHz to 300\,kHz in X2~\cite{zhang_frequency_2024} (see Methods). It is thus crucial to explore the origin of linewidth broadening in the crystal. We begin by studying the relationship between line b linewidth and temperature, recorded in Fig.~\ref{fig:fig2}a for C10, C13, and X2 (orange circle, green diamond, purple square). 
When multiple linescans are taken for a crystal at the same temperature (within \SI{1}{K}), the linewidth is represented by their weighted average. 
Across the measured temperatures, no temperature dependence is observed.

On the other hand, we observe clear linewidth dependence on the \thor{} doping concentration and the individual transition line. Figure~\ref{fig:fig2}b shows the FWHM linewidth as a function of \thor{} doping concentration for all three crystals. Filled markers correspond to line b while unfilled markers correspond to line c. Each marker represents the weighted average of data over all temperatures. Linear fits to line b [line c] yield a slope of 17.1(7) [37(3)]\,\unit{kHz/10^{18} cm^{-3}} and intercept of 24(2) [31(5)]\,\unit{kHz}. The weighted average of the measured linewidths of C10 line b is \SI{30(1)}{kHz}, and the narrowest feature we observe in C10 has an FWHM of 20(3)\,kHz.

While we have not directly measured the linewidth of our VUV comb, spectral analysis suggests it is on the order of 1\,kHz~\cite{zhang_tunable_2022}. Our nuclear spectroscopy measurements provide a further insight: the measured linear relationship of linewidth to near zero concentration suggests that the comb linewidth is below the observed values reported here.

\begin{figure*}[th!]
    \includegraphics[width=14cm]{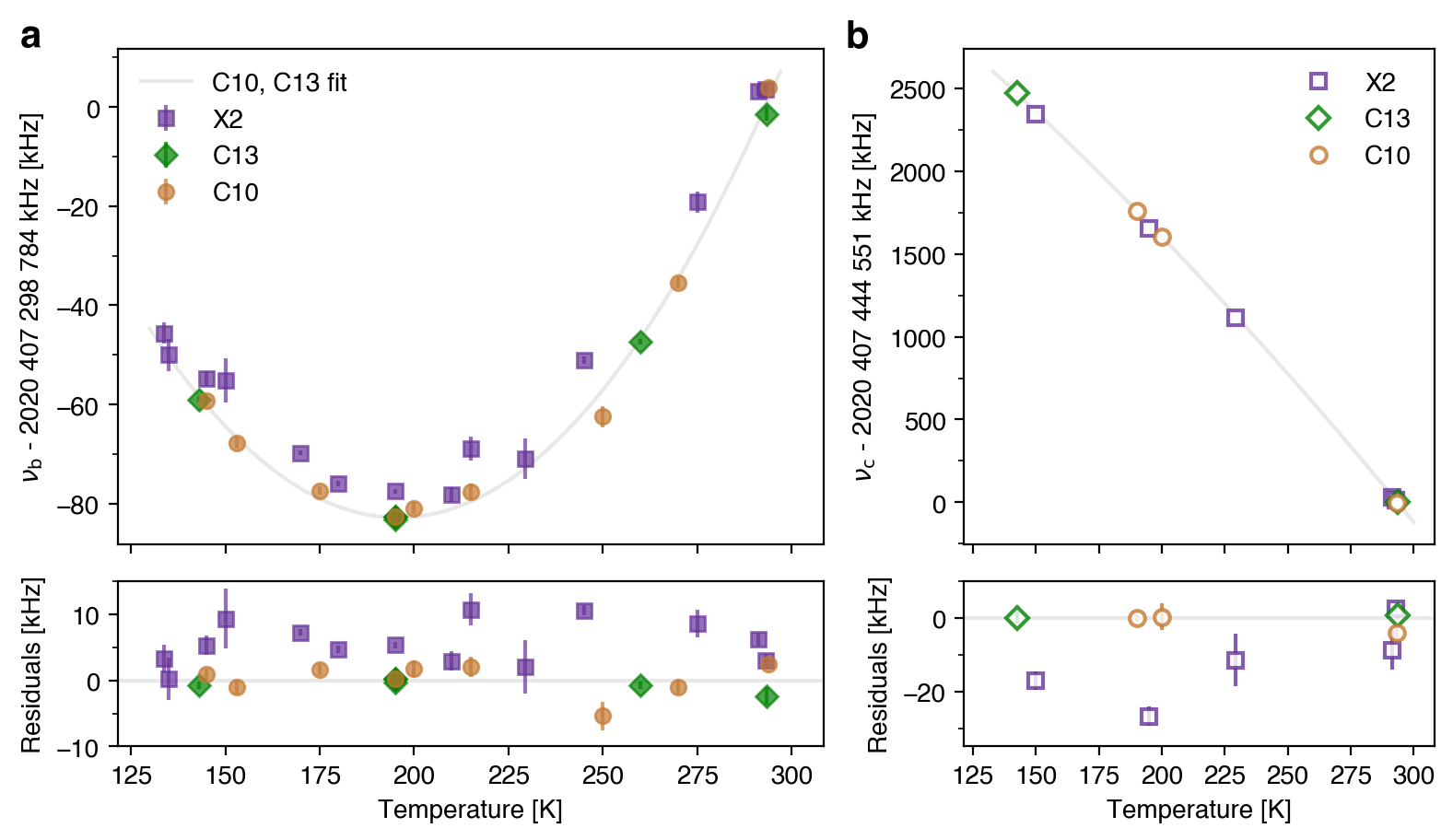}
    \caption{Temperature reproducibility of the nuclear clock. For both plots, the optical frequency of the nuclear clock transition is plotted against temperature, with error bars derived from the Lorentzian fit uncertainty. Multiple linescans taken at the same temperature within 1 K are represented by their weighted average. A quadratic fit to the C10 and C13 data is shown in light grey. Residuals from the quadratic fit are displayed in the bottom panel.
    \textbf{a}, Temperature dependence of line~b frequency $\nu_\mathrm{b}$. The minimum in this graph at 195(5)\,K indicates a temperature where  $\nu_\mathrm{b}$ has no 1st order frequency dependence on temperature, an ideal operating point for a clock. 
    \textbf{b}, Temperature dependence of line~c frequency $\nu_\mathrm{c}$, the most sensitive to temperature variation. This frequency can be used to precisely monitor the crystal temperature.}
    \label{fig:fig3}
\end{figure*}

\medskip
\noindent\textbf{Zero-shift temperature and co-thermometry}
\smallskip
\par\nobreak
We measure the nuclear transition center frequency at various crystal temperatures ranging from 134\,K to 293\,K. As shown in Fig.~\ref{fig:fig3}, the transition frequencies have good agreement among the three different crystals at all measured temperatures. For both line b and line c, the center frequencies $\nu_\mathrm{b}$ and $\nu_\mathrm{c}$ are plotted as a function of temperature. Points from the same crystal taken within 1\,K temperature are represented by their weighted average. The gray lines show the fit of a quadratic function to the C10 and C13 data, with the fit residuals plotted in the lower panels (see Extended Data Table \ref{mytable} for fitted parameters). 

Line b has the lowest temperature sensitivity across the four observed transition lines in Ref.~\cite{higgins_temperature_2025}, changing by only 80\,kHz over the measured temperature range. More importantly, we observe a local minimum of line b frequency at $T_\mathrm{0}=$ 195(5)\,K (Fig. \ref{fig:fig3}a), where the first-order dependence of the transition frequency on temperature vanishes. The X2 data shows a larger frequency spread compared to C10 and C13. We speculate that this may arise from stress-induced frequency shifts from the mounting glue contracting as the temperature decreases. The smaller crystal size may also make X2 more susceptible to local laser heating effects.

In contrast, line~c frequency has the strongest dependence on temperature and varies by 2.5\,MHz over the same range (Fig.~\ref{fig:fig3}b). Leveraging the greater temperature sensitivity, a co-thermometry setup can be implemented using the line~c frequency to directly measure the crystal temperature \textit{in-situ}. Assuming we can measure $\nu_\mathrm{c}$ to 1\,kHz uncertainty, this corresponds to a temperature uncertainty of 0.06\,K  at $T_0$. Thus, guided by line~c, the crystal temperature can be set to $T=T_\mathrm{0}\pm0.06\ \mathrm{K}$. Then, a 1\,mK temperature reproducibility via active stabilization would result in a line~b frequency reproducibility of 1\,mHz, corresponding to a fractional frequency reproducibility of $6\times 10^{-19}$. 

\medskip
\noindent\textbf{Frequency reproducibility over time and crystals}
\smallskip
\par\nobreak
Over a time period of one year, the nuclear transition frequency has shown good reproducibility at the kHz level among the three crystals. All measurements are referenced to the Sr clock laser~\cite{aeppli_clock_2024, oelker_demonstration_2019}, which enables consistent frequency comparisons over many months. Known systematic shifts, including the silicon reference cavity drift~\cite{milner_demonstration_2019}, are accounted for (see Methods).

\begin{figure}[th!]
    \includegraphics[width=\columnwidth]{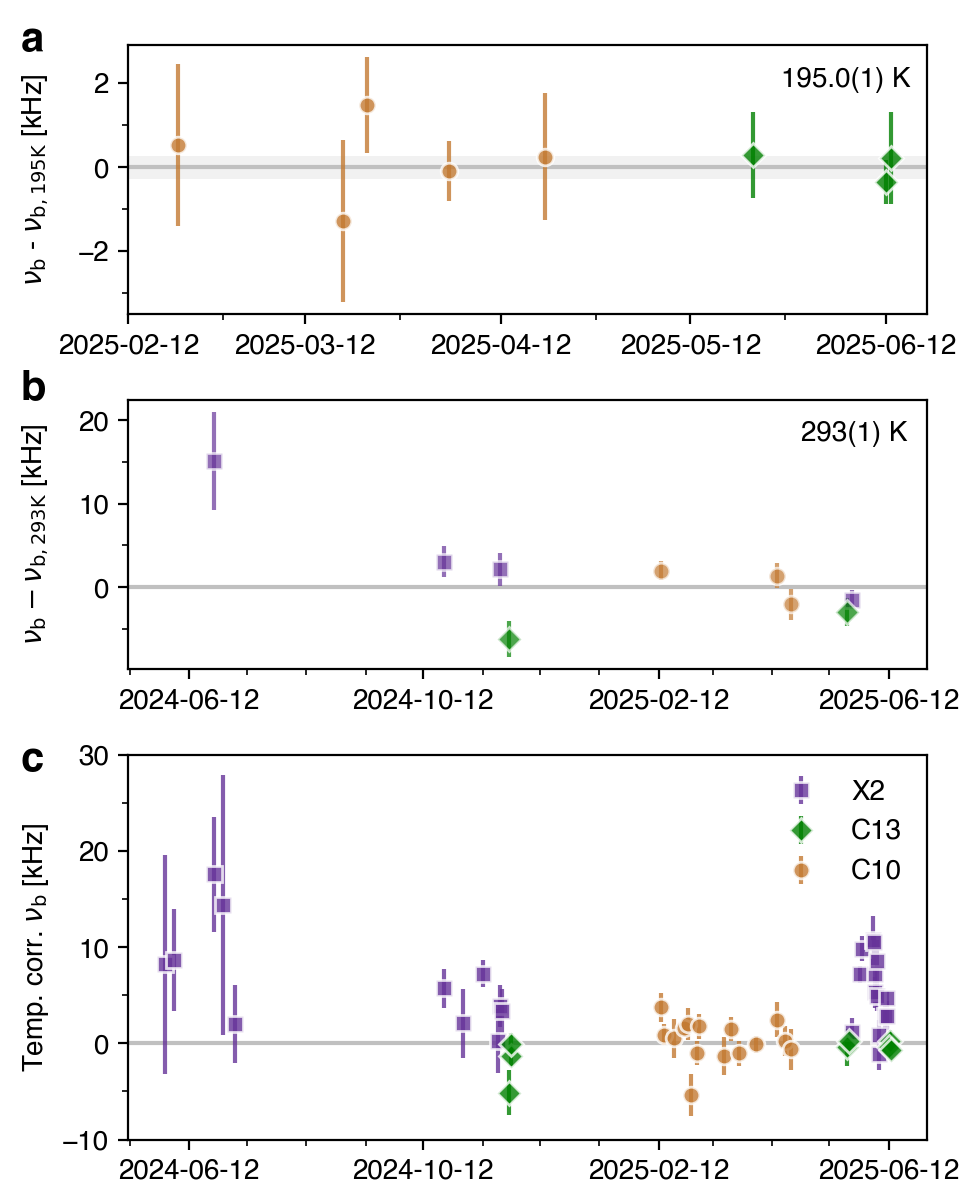}
    \caption{
       Time record of the line~b nuclear transition frequency across all three crystals.  Error bars from the Lorentzian fit are shown.
       \textbf{a}, Frequency $\nu_\mathrm{b}$ at the zero-shift temperature 195.0(1)\,K over a four month period, relative to the average frequency of line~b at 195\,K. The standard error, shown with light gray shading, is 280 Hz.
       \textbf{b}, Frequency $\nu_\mathrm{b}$ of line~b at 293(1)\,K over a one year period, relative to the average frequency of line~b at 293\,K. 
       \textbf{c}, Residuals of frequency $\nu_\mathrm{b}$ from the quadratic fit (Fig.~\ref{fig:fig3}b) shown against time. 
       By subtracting the temperature-induced shifts, all of our historical line b data are organized together. No systematic drift of the nuclear clock frequency, calibrated against the \sr{} optical clock, is observed. Some overlapping data points have been offset by a maximum of 4 days for clarity. Error bars include the uncertainty propagated from the temperature correction.
       }
       
    \label{fig:fig4}
\end{figure}

Figure~\ref{fig:fig4}a shows the frequency of line b at the zero-shift temperature, 195.0(1)\,K for C10 and C13 over a four month period. The weighted mean is $\nu_\mathrm{b,195K}=2,020,407,298,701.16$\,kHz, with a standard error of 0.28\,kHz (light gray band) and reduced chi-squared $\tilde{\chi}^2=0.4$. Error bars represent the 1-$\sigma$ uncertainty from the Lorentzian fit. 

In comparison, Fig.~\ref{fig:fig4}b shows the center frequency of line~b at 293(1)\,K in all three crystals over a one year time period. The weighted average is $\nu_\mathrm{b,293K}=2,020,407,298,787$\,kHz with standard error of 2 kHz. For some room temperature data, the crystal temperature is not actively stabilized, which could account for some frequency spread. 

To visualize the temporal consistency of the line~b data across all temperatures, we subtract the measured temperature-dependent shifts using the fit shown in Fig.~\ref{fig:fig3}. The result shown in Fig.~\ref{fig:fig4}c contains similar data as the residuals shown in the lower panel of Fig.~\ref{fig:fig3}a, but is plotted against date instead of temperature. Error bars represent the uncertainty from the Lorentzian fit and also the error from subtracting the temperature shift. Line scans from earlier dates have fewer points near the peak compared to the later dates, leading to a larger Lorentzian fit uncertainty. Across the three crystals over the one year period, the center frequencies show good agreement at the kilohertz level.

\medskip
\noindent\textbf{Inhomogeneous broadening mechanism}
\smallskip
\par\nobreak
Initial estimates for \thor:\caf\ \cite{kazakov_performance_2012} predicted that the spectroscopic linewidth would be on the order of 1\,kHz, dominated by the inhomogeneous broadening from the magnetic dipole interaction. Other mechanisms such as nuclear electric quadrupole interaction with phonons and second-order Doppler broadening were estimated to be hundreds of Hz. However, we observe much larger linewidths of 10s to 100s of kHz, suggesting a broadening mechanism different from previous considerations~\cite{rellergert_constraining_2010, kazakov_performance_2012}.

We suggest that the observed linewidth is dominated by crystal inhomogeneity due to microstrain from neighboring \thor\ defects that distort the local EFG. The density of thorium in the studied crystals ($0.3\times 10^{18}$~\unit{cm^{-3}} to $4\times 10^{18}$~\unit{cm^{-3}}) corresponds to an average distance of $d=$\,60\unit{\angstrom} to 150\unit{\angstrom} between \thor\ dopants. Although this distance is large compared to the \caf\ lattice constant of \SI{5.5}{\angstrom}, the \thor\ site distorts the crystal lattice, creating a long-range microstrain extending over many lattice spacings~\cite{eshelby_continuum_1956, flocken_asymptotic_1970}. As each defect is randomly dispersed throughout the crystal, we expect each \thor{} nucleus to see different $V_\mathrm{zz}$ and $\rho_\mathrm{e}(0)$ depending on the relative distance and orientation of nearby \thor{} doping sites. The nuclear transition frequency will be shifted positively or negatively depending on the orientation and position of surrounding defects. As the density of thorium increases, the microstrain from nearby thorium overlaps more, leading to the increased inhomogeneous broadening. 
This scenario has been described in theoretical~\cite{stoneham_shapes_1969} and experimental~\cite{pelzl_influence_1975, kawamura_quadrupolar_1956} nuclear quadrupole resonance studies of defect broadening in crystals, where a linear dependence of linewidth on concentration is expected assuming randomly distributed point defects. The non-zero intercept of approximately 25\,kHz may be attributed to the presence of known \caf{} crystal defects not connected with \thor{}. 

The observation that line~c is always broader than line b may be similarly understood. The temperature dependence shown in Fig.~\ref{fig:fig3} can be interpreted as a rough proxy for stress sensitivity, as increasing the crystal temperature or adding microstrain both distort the lattice, changing the electron density and electric field gradients. As $\nu_\mathrm{c}$ changes more with temperature than $\nu_\mathrm{b}$, the inhomogeneous broadening should be greater for line~c than line~b in the same defect concentration environment.  

Experimentally we find that the Lorentzian fit consistently yields smaller residuals than Gaussian lineshape fit. A Lorentzian lineshape is expected for transitions dominated by electric field gradient effects in low concentration point-defect-broadened crystals~\cite{stoneham_shapes_1969, kanert_influence_1969}. 
For a Poissonian defect distribution and for the frequency shift due to microstrain falling off as inverse of cubic distance, the resulting lineshape is Lorentzian~\cite{cohen_quadrupole_1957}. 

The microstrain-related broadening mechanism can be experimentally  investigated further in several ways. Crystals doped with both \thor{} and $^{232}\mathrm{Th}$ could allow exploration of different defect concentration environments, as $^{232}\mathrm{Th}$ is expected to create the same defect sites as \thor{} but does not possess the 148\,nm nuclear transition. With sufficient laser intensity, spectral hole-burning or spin-echo experiments can also be used to confirm the inhomogeneous broadening mechanism.

Techniques such as x-ray diffraction or atomic force microscopy can give insight to the crystalline structure, as has been done for $^{232}\mathrm{Th}$:\caf~\cite{gong_structures_2024, beeks_nuclear_2022}. X-ray absorption fine structure spectroscopy has provided better understanding of the defect structure and interstitial positions for \thor:\caf\ \cite{takatori_characterization_2025}, which could be used to form theoretical models of the strain defect.  Finally, we note that crystals where thorium is naturally part of the crystal lattice rather than a defect, such as $\mathrm{ThF_4}$~\cite{pastor_preparation_1974} or $\mathrm{Th(SO_4)_2}$~\cite{morgan_spinless_2025} should be free from this microstrain effect. In this case other defects intrinsic to the host material or due to \thor{} radioactivity may dominate.

\medskip
\noindent\textbf{Other systematics and outlook}
\smallskip
\par\nobreak
To build a practical nuclear clock, a number of other important systematic effects must be investigated. The interaction of the nuclear magnetic moment $\mu_I$ and a magnetic field $B$ results in the splitting of $\pm m$ levels, also known as the nuclear Zeeman effect. The shift of each level is given by $\mu_ImB$, corresponding to roughly 1 kHz/G depending on the specific transition and orientation of $B$ with respect to the EFG. The stray magnetic field in our current experiment, estimated to be $< 1$\,G, would not cause a significant effect in the \thor:\caf\ crystals, but this may become important for future crystals with less inhomogenous broadening. Additionally, the intensity and frequency of the Zeeman split lines can yield insight to the intrinsic magnetic environment~\cite{reissner_mossbauer_2021}. By changing the magnetic field orientation, the EFG tensor within the crystal can also be characterized~\cite{liechti_nmr-nqr_1989}. Another interesting investigation would be to subject the crystal to different stresses and measure potential frequency shifts as the electron density and EFG changes, which may clarify the observed X2 frequency scatter. 

The fractional frequency stability of a clock is characterized by
    $\sigma\propto \frac{\Delta f}{ f_0S}\sqrt{T_{\mathrm{cycle}}/\tau}$. 
Based on the current experimental conditions achieved for the C10 crystal, the linewidth $\Delta f=\ $\SI{30}{kHz}, center frequency $f_0=\ $ \SI{2020.407}{THz}, cycle time $T_\mathrm{cycle}=\ $ \SI{600}{s}, signal-to-noise ratio $S=20$ over $T_\mathrm{cycle}$,  and $\tau$ the averaging time~\cite{ludlow_optical_2015}, we expect $\sigma\approx 10^{-11}/\sqrt{\tau}$. While far below the performance of today's atomic clocks, this can be used as a guideline for future nuclear clock development, namely by improving $\Delta f$, $S$, and $T_{\mathrm{cycle}}$.

As previously discussed, in a crystal free of inhomogeneous microstrain broadening the linewidth may be reduced to the kHz~level, improving $\sigma$ by more than an order of magnitude. With a higher \thor\ doping concentration crystal or with a higher intensity excitation laser, the signal can be linearly improved. Suppressing the background counts via veto detection~\cite{hiraki_controlling_2024, hiraki_experimental_2024} or using a crystal with lower intrinsic background also increases the signal-to-noise ratio $S$. 

A huge gain can be achieved by reducing $T_\mathrm{cycle}$, which currently consists of 400\,s of laser irradiation and 200\,s of fluorescence photon detection. This could be obtained by using a higher power laser and improved readout techniques. The Rabi frequency for our current laser is estimated at approximately $0.1$\,Hz, much smaller than the observed linewidth of 30\,kHz. To enable coherent control of the nuclear population, the Rabi frequency should be on the order of the inhomogeneous linewidth or greater, which requires a substantial laser power upgrade or linewidth reduction. With a single-frequency continuous-wave laser rather than a frequency comb, the excited state could be read out through absorption instead of nuclear decay, potentially allowing continuous clock operation.

The coupling of electronic and nuclear states within the crystal could also be harnessed to control the nuclear population. A laser-induced quenching effect has been observed for doped \thor\ crystals, where shining off-resonant 200\,nm to 400\,nm light reduced the excited state lifetime by up to a factor of three~\cite{schaden_laser-induced_2024, terhune_photo-induced_2024}. A similar quenching effect was first reported with 29\,keV X-ray beams used to indirectly excite the nuclear transition~\cite{hiraki_controlling_2024, guan_x-ray_2025}. If these effects could yield higher signal over a shorter collection time, it would help improve $\sigma$. Similarly, if a nuclear state-dependent electronic transition exists in the crystal, it could be used to quickly readout the clock state with higher $S$ than the current fluorescence detection.

An important application of the nuclear clock to fundamental physics is constraining variations of fundamental constants, which are proposed in many ultra-light dark matter (ULDM) models. The nuclear clock has orders of magnitude enhanced sensitivity to this effect compared to atomic clocks since the ULDM couples dominantly to the nucleus (quantum chromodynamic sector)~\cite{flambaum_enhanced_2006, fuchs_implications_2024, beeks_fine-structure_2024, caputo_sensitivity_2024}. One analysis~\cite{fuchs_implications_2024} shows that a \thor\ nuclear clock with frequency uncertainty of 5\,kHz can already begin to put competitive bounds on the ULDM mass and coupling strength. The lineshape may also be of interest for constraining high frequency temporal variations. Using the strontium atomic clock as a reference, our current and future time record of the nuclear transition frequency can sensitively probe ULDM.

In summary, we report quantum-state resolved nuclear spectroscopy on three separately grown \thor:\caf\ crystals and verify the frequency reproducibility of the \thor\ nuclear transition over a one year period. Taking detailed temperature dependence data for line~b ($m=\pm5/2 \rightarrow m=\pm3/2$), we identify a temperature $T=195(5)$\,K where the first order temperature dependence of $\nu_\mathrm{b}$ vanishes. At this temperature we find the reproducibility of $\nu_\mathrm{b}$ is 280 Hz across two different crystals over four months. By performing a co-thermometry scheme with a more temperature sensitive transition such as line~c, the systematic temperature shift could be suppressed below the $10^{-18}$ fractional uncertainty level. We also observe a linear dependence of the nuclear transition linewidth on \thor\ doping concentration, which sheds light on how the \thor\ dopant distorts the crystalline host. This work shows the reliability of \thor:\caf\ crystals as a frequency reference and the first characterization of crystal-limited linewidths, representing a crucial step towards field-deployable solid-state nuclear clocks.

\noindent \textbf{Acknowledgments.} 
We thank K. Kim, A. Aeppli, W. Warfield, D. Lee, B. Lewis, and Z. Hu for running the Sr clock and ultrastable silicon cavity.
We thank  P. Li, K. Hagen, D. Warren, C. Schwadron, D. Reddy, and K. Li for technical assistance and M. Ashton, B. C. Denton and M. R. Statham for help in dealing with radioactive samples. 

We acknowledge funding support from National Science Foundation QLCI OMA-2016244, DOE quantum center of Quantum System Accelerator, Army Research Office (W911NF2010182), Air Force Office of Scientific Research (FA9550-19-1-0148), National Science Foundation PHY-2317149, and National Institute of Standards and Technology. Part of this work has been funded by the European Research Council (ERC) under the European Union’s Horizon 2020 research and innovation programme (Grant Agreement No. 856415) and the Austrian Science Fund (FWF) [Grant DOI: 10.55776/F1004, 10.55776/J4834, 10.55776/ PIN9526523]. The project 23FUN03 HIOC [Grant DOI: 10.13039/100019599] has received funding from the European Partnership on Metrology, co-financed from the European Union’s Horizon Europe Research and Innovation Program and by the Participating States. We thank the National Isotope Development Center of DoE and Oak Ridge National Laboratory for providing the Th-229 used in this work.

\smallskip
\noindent \textbf{Author contributions.} T.O., J.F.D., C.Z., J.S.H., and J.Y. performed the experiment and analysed data; K.B., T.Si., and T.Sc. grew the thorium-doped crystals and characterized its performance. All authors wrote the manuscript.

\smallskip
\noindent \textbf{Competing interests.} The authors declare no competing interests.

\smallskip
\noindent \textbf{Correspondence and requests for materials} should be addressed to Jun Ye.


\bibliography{references}


\bigskip
\noindent\textbf{\Large Methods}
\smallskip
\par\nobreak
\medskip
\noindent\textbf{Frequency comb and detection setup}
\smallskip
\par\nobreak
The spectroscopic data were taken using a vacuum ultraviolet frequency comb generated by cavity-enhanced high harmonic generation (HHG). The details of this procedure are described in previous publications \cite{jones_phase-coherent_2005, gohle_frequency_2005, zhang_tunable_2022}. Briefly, we generate a high power near infrared frequency comb with average power around 40\,W and a repetition frequency of 75\,MHz (IMRA America, center wavelength $\sim$1040\,nm). The comb is coupled to a femtosecond enhancement cavity for passive amplification of the comb power to 5\,kW, achieving high enough pulse power for HHG in xenon gas. We outcouple the seventh harmonic from the cavity (center frequency $\sim$148\,nm) and direct it onto the thorium sample. 

The fundamental comb light is locked to the JILA strontium clock laser at 698\,nm~\cite{aeppli_clock_2024}. This clock laser is locked to a Menlo frequency comb, which is stabilized to a cryogenic ultra-stable silicon cavity~\cite{oelker_demonstration_2019, matei_15_2017}. This allows for an absolute, frequency-based measurement of the thorium transition frequency. Additionally, the fundamental comb is locked to a near-IR non-planar ring oscillator (NPRO) at 1064\,nm, which itself is locked to the same Menlo comb. Locking the fundamental comb at two points in the optical domain surrounding the center wavelength provides further stabilization that may narrow the VUV comb linewidth compared to previous works. Extended Data Fig.~\ref{fig:locking} summarizes this new locking scheme. We note that in previous work, the frequency of the comb was swept about 100\,kHz during the irradiation window for each point of the line scan, which may distort the fitted linewidth compared to later line scans, where the laser frequency was kept fixed. However, the fitted line center will not be distorted. Consequently, line scans with frequency sweeping are excluded from Fig.~\ref{fig:fig2} but are included in Fig.~\ref{fig:fig3} and Fig.~\ref{fig:fig4}.

Our photon detection scheme has been described previously~\cite{zhang_frequency_2024}. We place the crystal into the focus of a parabolic mirror to collect photons radiated from the thorium nuclei. The light collimated from the parabolic mirror is spectrally filtered and imaged onto a photomultiplier tube (Hamamatsu). The line scans are conducted by stabilizing the carrier envelope offset frequency $f_\mathrm{CEO}$ of the fundamental comb to –8\,MHz while scanning the repetition rate $f_\mathrm{rep}$. For each scan point, $f_\mathrm{rep}$ remains fixed while the crystal is irradiated for 400\,s, followed by a 200\,s detection window. For each transition, the comb mode number determination was done in previous work~\cite{zhang_frequency_2024}, allowing for a single scan of each lineshape for absolute frequency determination.  

\medskip
\noindent\textbf{Data analysis}
\smallskip    
\par\nobreak
Each scan point signal is normalized by the average intensity of the excitation laser to reduce fluctuations from laser intensity drift. 
The excitation laser intensity is monitored using a camera capturing the visible fluorescence of the crystal under laser irradiation, similar to that presented in Ref.~\cite{tiedau_laser_2024}. This fluorescence, which arises from self-trapped exciton emission~\cite{stellmer_radioluminescence_2015}, depends strongly on the laser power and not on laser frequency. In some scans, we observed the background count rate after the scan was higher than before the scan started, distorting the lineshape. This is corrected for by subtracting a linear baseline before other corrections.

The long excitation life of the nuclear transition also presents a ``memory" effect and may distort the measured line shape. In previous publications this was mitigated by performing the frequency scan in forward and reverse directions~\cite{zhang_frequency_2024}. Using the measured lifetime of the \thor{} transition in \caf{}~\cite{tiedau_laser_2024, zhang_frequency_2024}, we now remove this memory effect in post-processing of the data similar to that performed in Ref.~\cite{tiedau_laser_2024} so that a bidirectional scan is not needed and the intrinsic lineshape is clearly presented. For a typical scan, we first take two points far from resonance to establish the baseline counts $B$. The exponential decay from the $i$th detection window can then be fitted to $y_i(t)=A_ie^{t/\tau}+B$, where $A_i$ is the signal amplitude and $\tau=641$\,s is the measured nuclear excited state lifetime in seconds~\cite{zhang_frequency_2024}. The exponential decay from the previous detection window can be extrapolated to the current one and then subtracted out. The memory corrected detection window is $y_{i,corr}(t)=y_i(t)-A_{i-1}e^{-(t+T)/\tau}$, where $T$ is time since the beginning of the $i-1$ detection window.

For data taken over many months, a slight frequency shift from slow drift of the silicon cavity needs to be accounted for. By fitting a line to the silicon cavity (Si3) absolute frequency~\cite{milner_demonstration_2019} measured against the JILA Sr1 atomic clock~\cite{aeppli_clock_2024} from May~2024 to March~2025, a drift of $-1.5$\,Hz/day is extracted. This corresponds to $-15.6$\,Hz/day for the thorium transition frequency, which accumulates to more than 5\,kHz over 12\,months. The shift is calculated for each date and subtracted from each fitted frequency. 

\renewcommand{\figurename}{Extended Data Fig.}
\renewcommand{\tablename}{Extended Data Table}

\medskip
\noindent\textbf{\Large Extended data}
\smallskip
\par\nobreak
\begin{table}[H]
\caption{Quadratic fit parameters for center frequency $\nu$ versus temperature $T$, where $\nu=\alpha T^2+\beta T + \gamma$.}
\label{mytable}

\begin{tabular}{l|ll}
                     & line b   & line c          \\ \hline
$\alpha$ [kHz/K$^2$] & 0.0088(2) & -0.0140(4) \\
$\beta$  [kHz/K]     & -3.44(7)  & -10.2(2)    \\
$\gamma$ [kHz]       & 254(7)  & 4210(20)  
\end{tabular}
\end{table}

\setcounter{figure}{0}
\begin{figure*}
    \includegraphics[width=18cm]{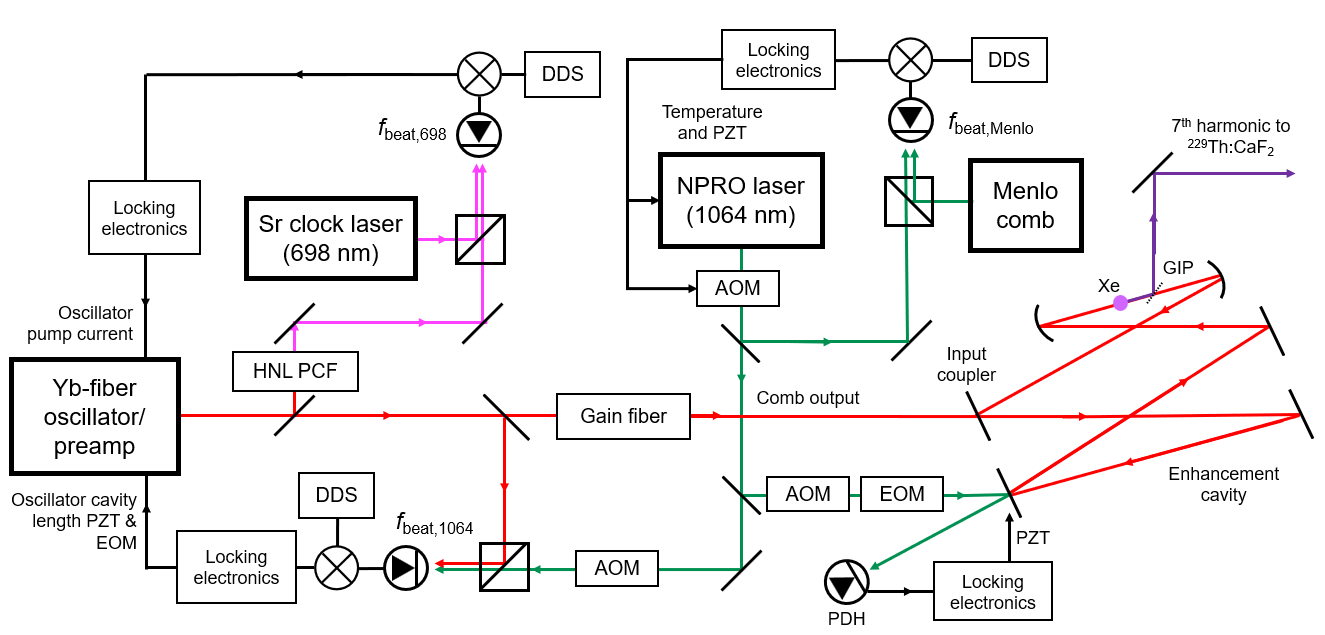}
    \caption{Frequency comb locking scheme for data taken after September 15, 2024. A Yb-fiber oscillator from IMRA America generates the IR frequency comb centered at $\sim 1040$\,nm, which is coupled into our enhancement cavity for VUV harmonic generation. The IR frequency comb is locked at two points, which are scanned to adjust the repetition rate $f_\mathrm{rep}$ while keeping the carrier envelope offset $f_\mathrm{CEO}=-8$\,MHz. For the first point, 698\,nm comb light generated from a highly nonlinear photonic crystal fiber (HNL PCF) beats with the Sr clock laser, indicated by $f_\mathrm{beat, 698}$. The desired beat frequency is set by a direct digital synthesizer (DDS), and the error signal feeds back onto the oscillator pump current. For the second point, 1064\,nm comb light beats with a non-planar ring oscillator (NPRO) Mephisto laser, indicated by $f_\mathrm{beat, 1064}$. The error signal feeds back onto the oscillator cavity length and electro-optic modulator (EOM). The NPRO laser itself is stabilized to the Menlo comb, indicated by $f_\mathrm{beat, Menlo}$, which feeds back onto the NPRO cavity length (PZT, piezo-electric transducer), temperature, and an acousto-optic modulator (AOM). The Sr clock laser is locked to the Menlo comb, and the Menlo comb is locked to the ultrastable silicon cavity (not shown). This scheme is designed to give the comb more robustness and frequency stability. NPRO light is used to stabilize the enhancement cavity using a Pound-Drever-Hall lock, with feedback onto one of the enhancement cavity mirrors. }
    \label{fig:locking}
\end{figure*}

\end{document}